\documentclass[11pt]{article}
\usepackage[left=1in, right=1in, top=1in, bottom=1in]{geometry}
\usepackage{CJK}
\usepackage{latexsym,bm}
\usepackage{xypic}
\usepackage{amsmath}
\usepackage{wasysym}
\usepackage{indentfirst}
\usepackage{amssymb}
\usepackage{dsfont}
\usepackage{amsthm}
\begin{document}
\newcommand{\fr}[2]{\frac{\;#1\;}{\;#2\;}}
\newtheorem{theorem}{Theorem}[section]
\newtheorem{lemma}{Lemma}[section]
\newtheorem{proposition}{Proposition}[section]
\newtheorem{corollary}{Corollary}[section]
\newtheorem{conjecture}{Conjecture}[section]
\newtheorem{remark}{Remark}[section]
\newtheorem{definition}{Definition}[section]
\newtheorem{example}{Example}[section]
\newtheorem{notation}{Notation}[section]
\numberwithin{equation}{section}
\newcommand{\Aut}{\mathrm{Aut}\,}
\newcommand{\CSupp}{\mathrm{CSupp}\,}
\newcommand{\Supp}{\mathrm{Supp}\,}
\newcommand{\rank}{\mathrm{rank}\,}
\newcommand{\col}{\mathrm{col}\,}
\newcommand{\len}{\mathrm{len}\,}
\newcommand{\leftlen}{\mathrm{leftlen}\,}
\newcommand{\rightlen}{\mathrm{rightlen}\,}
\newcommand{\length}{\mathrm{length}\,}
\newcommand{\bin}{\mathrm{bin}\,}
\newcommand{\wt}{\mathrm{wt}\,}
\newcommand{\Wt}{\mathrm{Wt}\,}
\newcommand{\diff}{\mathrm{diff}\,}
\newcommand{\lcm}{\mathrm{lcm}\,}
\newcommand{\GL}{\mathrm{GL}\,}
\newcommand{\SJ}{\mathrm{SJ}\,}
\newcommand{\LG}{\mathrm{LG}\,}
\newcommand{\bij}{\mathrm{bij}\,}
\newcommand{\dom}{\mathrm{dom}\,}
\newcommand{\fun}{\mathrm{fun}\,}
\newcommand{\SUPP}{\mathrm{SUPP}\,}
\newcommand{\supp}{\mathrm{supp}\,}
\newcommand{\End}{\mathrm{End}\,}
\newcommand{\Hom}{\mathrm{Hom}\,}
\newcommand{\ran}{\mathrm{ran}\,}
\newcommand{\row}{\mathrm{row}\,}
\newcommand{\Mat}{\mathrm{Mat}\,}
\newcommand{\rk}{\mathrm{rk}\,}
\newcommand{\rs}{\mathrm{rs}\,}
\newcommand{\piv}{\mathrm{piv}\,}
\newcommand{\perm}{\mathrm{perm}\,}
\newcommand{\rsupp}{\mathrm{rsupp}\,}
\newcommand{\inv}{\mathrm{inv}\,}
\newcommand{\orb}{\mathrm{orb}\,}
\newcommand{\id}{\mathrm{id}\,}
\newcommand{\soc}{\mathrm{soc}\,}
\newcommand{\unit}{\mathrm{unit}\,}
\newcommand{\word}{\mathrm{word}\,}
\newcommand{\Sym}{\mathrm{Sym}\,}

\renewcommand{\thefootnote}{\fnsymbol{footnote}}

\title{An Elementary Approach to MacWilliams Extension Property and Constant Weight Code with Respect to Weighted Hamming Metric}
\author{Yang Xu$^1$ \,\,\,\,\,\, Haibin Kan$^2$\,\,\,\,\,\,Guangyue Han$^3$}

\maketitle

\renewcommand{\thefootnote}{\fnsymbol{footnote}}

% 单位、地址、基金
\footnotetext{\hspace*{-6mm} \begin{tabular}{@{}r@{}p{16cm}@{}}
$^1$ & Shanghai Key Laboratory of Intelligent Information Processing, School of Computer Science, Fudan University,
Shanghai 200433, China.\\
&Shanghai Engineering Research Center of Blockchain, Shanghai 200433, China. {E-mail:xuyyang@fudan.edu.cn}\\
$^2$ & Shanghai Key Laboratory of Intelligent Information Processing, School of Computer Science, Fudan University,
Shanghai 200433, China.\\
&Shanghai Engineering Research Center of Blockchain, Shanghai 200433, China.\\
&Yiwu Research Institute of Fudan University, Yiwu City, Zhejiang 322000, China. {E-mail:hbkan@fudan.edu.cn} \\
$^3$ & Department of Mathematics, Faculty of Science, The University of Hong Kong, Pokfulam Road, Hong Kong, China. {E-mail:ghan@hku.hk} \\
\end{tabular}}

\vskip 3mm

{\hspace*{-6mm}{\bf Abstract}---In this paper, we characterize the MacWilliams extension property (MEP) and constant weight codes with respect to $\omega$-weight defined on $\mathbb{F}^{\Omega}$ via an elementary approach, where $\mathbb{F}$ is a finite field, $\Omega$ is a finite set, and $\omega:\Omega\longrightarrow\mathbb{R}^{+}$
is a weight function. Our approach relies solely on elementary linear algebra and two key identities for $\omega$-weight of subspaces derived from a double-counting argument. When $\omega$ is the constant $1$ map, our results recover two well-known results for Hamming metric code: (1) any Hamming weight preserving map between linear codes extends to a Hamming weight isometry of the entire ambient space; and (2) any constant weight Hamming metric code is a repetition of the dual of Hamming code.

\section{Introduction}
In 1962, MacWilliams proved in \cite{7} that any Hamming weight preserving map between linear codes extends to a Hamming weight isometry of the entire ambient space. This property is henceforth referred to as the MacWilliams extension property (MEP). In \cite{4}, Bogart, Goldberg and Gordon gave an elementary combinatorial proof of MEP by using binary matrices induced by $1$-dimensional subspaces. In \cite{11}, Ward and Wood gave another proof of MEP through the lens of character theory for finite abelian groups; moreover, using MEP, they provided another proof of the well-known result of Bonisoli that any constant weight Hamming metric code is a  repetition of the dual of Hamming codes (see \cite{3}). The latter result was also proved by Ward in \cite{10} by using group characters, and extended by Liu and Chen in \cite{5} by using value functions.

In this paper, we characterize MEP and constant weight code with respect to weighted Hamming metric, a generalization of Hamming metric which is determined by a weight function defined on the coordinate set (see \cite{2} or Section 2.1 for more details). Our approach, which avoids the use of group characters, is elementary in the sense that it relies solely on elementary linear algebra and two identities for $\omega$-weight of subspaces.

The rest of the paper is organized as follows.

In Section 2.1, we introduce necessary preliminaries, including some notation and definitions for weighted Hamming metric, MEP and constant weight code. In Section 2.2, we state our results on MEP. We start with two identities for $\omega$-weight of subspaces (Proposition 2.1), which are derived from a double-counting argument with the help of $q$-binomial coefficient (see Section 3 for more details). Next, we give necessary and sufficient conditions for local equivalence, and give a sufficient condition for local equivalence to imply global equivalence (Theorems 2.1 and 2.2). Finally, we derive two necessary and sufficient conditions for MEP (Theorem 2.3). In Section 2.3, we give a necessary and sufficient condition for constant weight code (Proposition 2.2), and then characterize constant weight code in terms of generator matrices (Theorem 2.4). In Section 3, we give detailed proofs of all the results stated in Section 2.

\section{Preliminaries}

\setlength{\parindent}{2em}
Throughout the paper, let $\mathbb{F}$ be a finite field with $|\mathbb{F}|=q$, $\Omega$ be a nonempty finite set, and let $\mathbf{H}\triangleq\mathbb{F}^{\Omega}$. Any $\mathbb{F}$-subspace $C\leqslant_{\mathbb{F}}\mathbf{H}$ is referred to as a \textit{code}. For $\alpha\in\mathbf{H}$ and $A\subseteq\mathbf{H}$, the \textit{Hamming support of $\alpha$ and $A$} are respectively defined as
\begin{equation}\supp(\alpha)\triangleq\{i\in\Omega\mid\alpha_i\neq0\},\end{equation}
\begin{equation}\chi(A)\triangleq\{i\in\Omega\mid\exists~\beta\in A~s.t.~\beta_i\neq0\}.\end{equation}
For $k\in\mathbb{Z}^{+}$, let $\mathbb{F}^{k}$ and $\mathbb{F}^{[k]}$ denote the sets of all the row vectors and column vectors over $\mathbb{F}$ of length $k$, respectively, and let $\Mat_{k,\Omega}(\mathbb{F})$ denote the set of all the matrices over $\mathbb{F}$ whose rows are indexed by $\{1,\dots,k\}$ and columns are indexed by $\Omega$. For $k\in\mathbb{Z}^{+}$ and $A\subseteq \mathbb{F}^{k}$, let
\begin{equation}A^{\bot}\triangleq\{\beta\in \mathbb{F}^{[k]}\mid \mbox{$\sum_{i=1}^{k}\alpha_i\beta_i=0$ for all $\alpha\in A$}\}.\end{equation}
For $k\in\mathbb{Z}^{+}$ and $G\in\Mat_{k,\Omega}(\mathbb{F})$, the \textit{column map of $G$} is the map $\tau:\Omega\longrightarrow\mathbb{F}^{[k]}$ defined as
\begin{equation}\forall~i\in\Omega:\text{$\tau(i)$ is equal to the $i$-th column of $G$}.\end{equation}

\subsection{Weighted Hamming metric}

From now on, we fix $\omega:\Omega\longrightarrow\mathbb{R}^{+}$. For $\beta\in\mathbf{H}$ and $A\subseteq\mathbf{H}$, the $\omega$-weight of $\beta$ and $A$ are respectively defined as
\begin{equation}\wt(\beta)=\sum_{i\in\supp(\beta)}\omega(i),\end{equation}
\begin{equation}\Wt(A)=\sum_{i\in\chi(A)}\omega(i).\end{equation}
One can verify that $d_{wH}:\mathbf{H}\times\mathbf{H}\longrightarrow\mathbb{R}$ defined as
$$d_{wH}(\alpha,\beta)=\wt(\beta-\alpha)$$
induces a metric on $\mathbf{H}$, which will henceforth be referred to as \textit{weighted Hamming metric} (see \cite{2}). When $\omega(i)=1$ for all $i\in\Omega$, weighted Hamming metric boils down to Hamming metric. In general, since the weight function $\omega$ takes values on each coordinate position, weighted Hamming metric can
be useful to model some specific kind of channels for which
the error probability depends on a codeword position (i.e., the
distribution of errors is nonuniform). Moreover, weighted Hamming metric agrees with channels in which it is necessary to correct all errors from a certain
set containing vectors with various configurations and weights (see \cite{2} and references therein for more details).

\setlength{\parindent}{2em}
Now, we give some definitions related to $\omega$-weight. We begin with the definitions of weight isometry, MEP and transitivity of $\mathbf{H}$.

\setlength{\parindent}{0em}
\begin{definition}
{\bf{(1)}}\,\,For $C\leqslant_{\mathbb{F}}\mathbf{H}$ and $f\in\Hom_{\mathbb{F}}(C,\mathbf{H})$, we say that $f$ preserves $\omega$-weight if $\wt(f(\alpha))=\wt(\alpha)$ for all $\alpha\in C$. An $\mathbb{F}$-automorphism $\varphi\in\End_{\mathbb{F}}(\mathbf{H})$ which preserves $\omega$-weight is referred to as an \textit{$\omega$-weight isometry}.

{\bf{(2)}}\,\,We say that $\mathbf{H}$ satisfies MEP with respect to $\omega$-weight if for any $C\leqslant_{\mathbb{F}}\mathbf{H}$ and $f\in\Hom_{\mathbb{F}}(C,\mathbf{H})$ such that $f$ preserves $\omega$-weight, there exists an $\omega$-weight isometry $\varphi\in\End_{\mathbb{F}}(\mathbf{H})$ such that $\varphi\mid_{C}=f$.

{\bf{(3)}}\,\,We say that $\mathbf{H}$ is transitive with respect to $\omega$-weight if for any $\alpha,\beta\in\mathbf{H}$ with $\wt(\alpha)=\wt(\beta)$, there exists an $\omega$-weight isometry $\varphi\in\End_{\mathbb{F}}(\mathbf{H})$ such that $\beta=\varphi(\alpha)$.
\end{definition}

\setlength{\parindent}{2em}
Next, following the spirits in \cite{1,8}, we define local and global equivalences between linear maps.

\setlength{\parindent}{0em}
\begin{definition}
Let $X$ be a finite dimensional $\mathbb{F}$-vector space, and let $f,g\in\Hom_{\mathbb{F}}(X,\mathbf{H})$. We say that $f$ and $g$ are locally $\omega$-equivalent if $\wt(f(\theta))=\wt(g(\theta))$ for all $\theta\in X$, and we say that $f$ and $g$ are globally $\omega$-equivalent if $g=\varphi\circ f$ for some $\omega$-weight isometry $\varphi\in\End_{\mathbb{F}}(\mathbf{H})$.
\end{definition}

\setlength{\parindent}{2em}
Local and global equivalences are closely related to MEP. Indeed, for $C\leqslant_{\mathbb{F}}\mathbf{H}$ and $f\in\Hom_{\mathbb{F}}(C,\mathbf{H})$, $f$ preserves $\omega$-weight if and only if $f$ and $\id_C$ are locally $\omega$-equivalent, and $f$ extends to an $\omega$-weight isometry if and only if $f$ and $\id_C$ are globally $\omega$-equivalent. Apparently, global equivalence always implies local equivalence; moreover, MEP is satisfied if and only if local equivalence always implies global equivalence.

\setlength{\parindent}{2em}
Now we define constant weight code.

\begin{definition}
For $C\leqslant_{\mathbb{F}}\mathbf{H}$, we say that $C$ is a constant weight code with respect to $\omega$ if $\wt(\alpha)=\wt(\beta)$ for all $\alpha,\beta\in C-\{0\}$.
\end{definition}

We end this subsection with the definition of unique decomposition property (UDP), which will be used for studying both MEP and constant weight code.

\begin{definition}
For $H,K\subseteq\Omega$, we say that $(H,K,\omega)$ satisfies the unique decomposition property (UDP) if for any $I\subseteq H$, $J\subseteq K$ with $\sum_{i\in I}\omega(i)=\sum_{j\in J}\omega(j)$, it holds that
$$\forall~b\in\mathbb{R}:|\{i\in I\mid \omega(i)=b\}|=|\{j\in J\mid \omega(j)=b\}|.$$
For $H\subseteq\Omega$, we simply say that $(H,\omega)$ satisfies UDP if $(H,H,\omega)$ satisfies UDP.
\end{definition}

\subsection{Results on MEP}

\setlength{\parindent}{2em}
Throughout this subsection, let $X$ be a finite dimensional $\mathbb{F}$-vector space with $\dim_{\mathbb{F}}(X)=k\geqslant1$, and let $f,g\in\Hom_{\mathbb{F}}(X,\mathbf{H})$. We begin by recalling the notion of \textit{$q$-binomial coefficient}. More precisely, for any $(n,r)\in\mathbb{N}^{2}$, following \cite{9}, the $q$-binomial coefficient of $(n,r)$ is defined as
\begin{eqnarray*}\begin{bmatrix}n\\r\end{bmatrix}\triangleq\begin{cases}
0,&n+1\leqslant r;\\
\prod_{i=1}^{r}\frac{q^{i+n-r}-1}{q^{i}-1},&n\geqslant r.
\end{cases}
\end{eqnarray*}
It is well-known that for $A\leqslant_\mathbb{F}X$ with $\dim_{\mathbb{F}}(A)=a$ and $b\in\mathbb{N}$ with $a\leqslant b$, the number of all the $b$-dimensional $\mathbb{F}$-subspaces of $X$ containing $A$ is equal to the $q$-binomial coefficient of $(k-a,b-a)$.

Our starting point is the following proposition.

\begin{proposition}
For $B\leqslant_{\mathbb{F}}X$ with $\dim_{\mathbb{F}}(B)=m$, it holds that
\begin{equation}\sum_{\theta\in B}\wt(f(\theta))=(q^{m}-q^{m-1})\Wt(f[B]).\end{equation}
Moreover, let $a\in\{0,1,\dots,k-1\}$, $A\leqslant_{\mathbb{F}}X$ with $\dim_{\mathbb{F}}(A)=a$, and let $m\in\{a+1,\dots,k\}$. Then, it holds that
\begin{equation}\sum_{(B\leqslant_{\mathbb{F}}X,\dim_{\mathbb{F}}(B)=m,A\subseteq B)}\Wt(f[B])=q^{k-m}\begin{bmatrix}k-a-1\\m-a-1\end{bmatrix}\Wt(f[X])+\begin{bmatrix}k-a-1\\m-a\end{bmatrix}\Wt(f[A]).\end{equation}
\end{proposition}

Proposition 2.1 leads to the following necessary and sufficient conditions for local equivalence.

\setlength{\parindent}{0em}
\begin{theorem}
{\bf{(1)}}\,\,If $f$ and $g$ are locally $\omega$-equivalent, then $\Wt(f[B])=\Wt(g[B])$ for all $B\leqslant_{\mathbb{F}}X$. Conversely, if there exists $m\in\{1,\dots,k-1\}$ such that $\Wt(f[B])=\Wt(g[B])$ for all $B\leqslant_{\mathbb{F}}X$ with $\dim_{\mathbb{F}}(B)=m$, then $f$ and $g$ are locally $\omega$-equivalent.

{\bf{(2)}}\,\,Let $(x_1,\dots,x_k)$ be an $\mathbb{F}$-basis of $X$, and let $L,M\in\Mat_{k,\Omega}(\mathbb{F})$ such that $f(\sum_{i=1}^{k}\gamma_ix_i)=\gamma L$, $g(\sum_{i=1}^{k}\gamma_ix_i)=\gamma M$ for all $\gamma\in\mathbb{F}^{k}$; moreover, let $\tau,\eta:\Omega\longrightarrow\mathbb{F}^{[k]}$ denote the column maps of $L$ and $M$, respectively. Then, $f$ and $g$ are locally $\omega$-equivalent if and only if for any $I\leqslant_{\mathbb{F}}\mathbb{F}^{[k]}$ with $\dim_{\mathbb{F}}(I)=1$, it holds that
\begin{equation}\sum_{(i\in\chi(f[X]),\tau(i)\in I)}\omega(i)=\sum_{(i\in\chi(g[X]),\eta(i)\in I)}\omega(i).\end{equation}
\end{theorem}

\setlength{\parindent}{2em}
Next, we show that local equivalence implies global equivalence when suitable UDP condition is satisfied.

\setlength{\parindent}{0em}
\begin{theorem}
If $f$ and $g$ are locally $\omega$-equivalent and $(\chi(f[X]),\chi(g[X]),\omega)$ satisfies UDP, then $f$ and $g$ are globally $\omega$-equivalent.
\end{theorem}

\setlength{\parindent}{2em}
Theorem 2.2 further leads to the following necessary and sufficient conditions for MEP.

\setlength{\parindent}{0em}
\begin{theorem}
$\mathbf{H}$ satisfies MEP with respect to $\omega$-weight if and only if $\mathbf{H}$ is transitive with respect to $\omega$-weight, if and only if $(\Omega,\omega)$ satisfies UDP.
\end{theorem}

\begin{remark}
{\bf{(1)}}\,\,When $\omega(i)=1$ for all $i\in\Omega$, the first part of Theorem 2.1 immediately follows from MEP for Hamming metric. For general $\omega$-weight, MEP may not hold, but Theorem 2.1 is always valid and therefore might be of independent interest.

{\bf{(2)}}\,\,Theorem 2.2 is of independent interest in the sense that we consider two maps $f$ and $g$ instead of the entire MEP condition for $\mathbf{H}$, and hence only a ``local'' UDP condition is required.

{\bf{(3)}}\,\,Theorem 2.3 is a special case of [12, Theorem VI.1], and when $q=2$, $(1)\Longleftrightarrow(3)$ of Theorem 2.3 is a special case of [6, Theorem 8]. The proof we provide here is simpler and more direct. When $\omega(i)=1$ for all $i\in\Omega$, UDP is always satisfied, and hence both Theorem 2.2 and Theorem 2.3 recover the MEP for Hamming metric (see \cite{4,7}).
\end{remark}

\subsection{Results on constant weight code}

\setlength{\parindent}{2em}
Throughout this subsection, let $C\leqslant_{\mathbb{F}}\mathbf{H}$ with $\dim_{\mathbb{F}}(C)=k\geqslant1$. We first derive the following necessary and sufficient condition for $C$ to be a constant weight code (with respect to $\omega$).

\setlength{\parindent}{0em}
\begin{proposition}
If $C$ is a constant weight code, then $\Wt(D)=\Wt(Q)$ for all $D,Q\leqslant_{\mathbb{F}}C$ with $\dim_{\mathbb{F}}(Q)=\dim_{\mathbb{F}}(D)$. Conversely, if there exists $m\in\{1,\dots,k-1\}$ such that all the $m$-dimensional $\mathbb{F}$-subspaces of $C$ have the same $\omega$-weight, then $C$ is a constant weight code.
\end{proposition}

\setlength{\parindent}{2em}
Next, we characterize constant weight code in terms of generator matrices.

\setlength{\parindent}{0em}
\begin{theorem}
Let $G\in\Mat_{k,\Omega}(\mathbb{F})$ satisfy that $C=\{\gamma G\mid\gamma\in\mathbb{F}^{k}\}$, i.e., $G$ is a generator matrix of $C$, and let $\tau:\Omega\longrightarrow\mathbb{F}^{[k]}$ be the column map of $G$. Then, the following two statements hold:

{\bf{(1)}}\,\,$C$ is a constant weight code if and only if there exists $\sigma\in\mathbb{R}$ such that
\begin{equation}\forall~I\leqslant_{\mathbb{F}}\mathbb{F}^{[k]}~s.t.~\dim_{\mathbb{F}}(I)=1:\sum_{(i\in\chi(C),\tau(i)\in I)}\omega(i)=\sigma.\end{equation}
Moreover, if (2.10) holds, then for any $D\leqslant_{\mathbb{F}}C$ with $\dim_{\mathbb{F}}(D)=s$, it holds that
\begin{equation}\Wt(D)=\frac{(q^{k}-q^{k-s})\sigma}{q-1};\end{equation}
{\bf{(2)}}\,\,If for any $I,J\leqslant_{\mathbb{F}}\mathbb{F}^{[k]}$ with $\dim_{\mathbb{F}}(I)=\dim_{\mathbb{F}}(J)=1$, it holds that
\begin{equation}\forall~b\in\mathbb{R}:|\{i\in\chi(C)\mid \tau(i)\in I,\omega(i)=b\}|=|\{j\in\chi(C)\mid \tau(j)\in J,\omega(j)=b\}|,\end{equation}
then $C$ is a constant weight code. Conversely, if $(\chi(C),\omega)$ satisfies UDP and $C$ is a constant weight code, then (2.12) holds true for all $I,J\leqslant_{\mathbb{F}}\mathbb{F}^{[k]}$ with $\dim_{\mathbb{F}}(I)=\dim_{\mathbb{F}}(J)=1$.
\end{theorem}

\setlength{\parindent}{2em}
Proposition 2.2 and Theorem 2.4 generalize counterpart results established for Hamming metric code, as detailed in the following remark.

\begin{remark}
When $\omega(i)=1$ for all $i\in\Omega$, Proposition 2.2 and Theorem 2.4 recover [5, Theorem 1] and the well-known result that constant weight Hamming metric codes are repetitions of the dual of Hamming codes (see \cite{3,5,10,11}). In Section 3, we will give a short proof of Proposition 2.2 by using Theorem 2.1, and then establish Theorem 2.4 with the help of Proposition 2.2.
\end{remark}

\section{Proofs of the results}
\setlength{\parindent}{2em}
In this section, we prove all the results stated in Section 2. We begin by establishing Proposition 2.1 by using a double-counting argument and $q$-binomial coefficient.

\begin{proof}[Proof of Proposition 2.1]
First, we prove (2.7). Let $D=f[B]$, and write $\dim_{\mathbb{F}}(D)=r$. Noticing that $|\{\theta\in B\mid f(\theta)=\beta\}|=q^{m-r}$ for all $\beta\in f[B]$, we have
\begin{eqnarray*}
\begin{split}
\sum_{\theta\in B}\wt(f(\theta))&=q^{m-r}\sum_{\beta\in D}\sum_{i\in\supp(\beta)}\omega(i)\\
&=q^{m-r}\sum_{i\in\Omega}|\{\beta\in D\mid\beta_i\neq0\}|\omega(i)\\
&=q^{m-r}\sum_{i\in\chi(D)}(q^{r}-q^{r-1})\omega(i)\\
&=(q^{m}-q^{m-1})\Wt(D),
\end{split}
\end{eqnarray*}
as desired. Next, we prove (2.8). Let $v$ denote the left hand side of (2.8). By (2.7), we have
\begin{eqnarray*}
\begin{split}
(q^{m}-q^{m-1})v&=\sum_{(B\leqslant_{\mathbb{F}}X,\dim_{\mathbb{F}}(B)=m,A\subseteq B)}\sum_{\theta\in B-\{0\}}\wt(f(\theta))\\
&=\sum_{\theta\in X-\{0\}}|\{B\leqslant_{\mathbb{F}}X\mid\dim_{\mathbb{F}}(B)=m,A\subseteq B,\theta\in B\}|\wt(f(\theta))\\
&=\left(\sum_{\theta\in A-\{0\}}\begin{bmatrix}k-a\\m-a\end{bmatrix}\wt(f(\theta))\right)+\left(\sum_{\theta\in X-A}\begin{bmatrix}k-(a+1)\\m-(a+1)\end{bmatrix}\wt(f(\theta))\right)\\
&=\begin{bmatrix}k-a-1\\m-a-1\end{bmatrix}\left(\sum_{\theta\in X}\wt(f(\theta))\right)+\left(\begin{bmatrix}k-a\\m-a\end{bmatrix}-\begin{bmatrix}k-a-1\\m-a-1\end{bmatrix}\right)\left(\sum_{\theta\in A}\wt(f(\theta))\right)\\
&=\begin{bmatrix}k-a-1\\m-a-1\end{bmatrix}(q^{k}-q^{k-1})\Wt(f[X])+q^{m-a}\begin{bmatrix}k-a-1\\m-a\end{bmatrix}(q^{a}-q^{a-1})\Wt(f[A]),
\end{split}
\end{eqnarray*}
which establishes (2.8), as desired.
\end{proof}

Next, we prove Theorem 2.1 with the help of Proposition 2.1.

\setlength{\parindent}{0em}
\begin{proof}[Proof of Theorem 2.1]
{\bf{(1)}}\,\,First, if $f$ and $g$ are locally $\omega$-equivalent, then for $B\leqslant_{\mathbb{F}}X$, by (2.7), we have
$$|B|(1-q^{-1})\Wt(f[B])=\sum_{\theta\in B}\wt(f(\theta))=\sum_{\theta\in B}\wt(g(\theta))=|B|(1-q^{-1})\Wt(g[B]),$$
which implies that $\Wt(f[B])=\Wt(g[B])$, establishing the first assertion. Next, we prove the ``conversely'' part. Noticing that
$$\sum_{(B\leqslant_{\mathbb{F}}X,\dim_{\mathbb{F}}(B)=m)}\Wt(f[B])=\sum_{(B\leqslant_{\mathbb{F}}X,\dim_{\mathbb{F}}(B)=m)}\Wt(g[B]),$$
an application of (2.8) to $a=0$ and $A=\{0\}$ implies that $\Wt(f[X])=\Wt(g[X])$. Without loss of generality, we assume that $m\geqslant2$. For an arbitrary $A\leqslant_{\mathbb{F}}X$ with $\dim_{\mathbb{F}}(A)=1$, we have
$$\sum_{(B\leqslant_{\mathbb{F}}X,\dim_{\mathbb{F}}(B)=m,A\subseteq B)}\Wt(f[B])=\sum_{(B\leqslant_{\mathbb{F}}X,\dim_{\mathbb{F}}(B)=m,A\subseteq B)}\Wt(g[B]),$$
which, together with $\Wt(f[X])=\Wt(g[X])$, $2\leqslant m\leqslant k-1$ and (2.8), implies that $\Wt(f[A])=\Wt(g[A])$, as desired.

{\bf{(2)}}\,\,We begin by noting that $\chi(f[X])=\{i\in\Omega\mid\tau(i)\neq0\}$, $\chi(g[X])=\{i\in\Omega\mid\eta(i)\neq0\}$. Moreover, for $U\leqslant_{\mathbb{F}}\mathbb{F}^{k}$ and $B\triangleq\{\sum_{i=1}^{k}\gamma_ix_i\mid\gamma\in U\}$, from $f[B]=\{\gamma L\mid\gamma\in U\}$, $g[B]=\{\gamma M\mid\gamma\in U\}$, we deduce that $\chi(f[B])=\{i\in\Omega\mid \tau(i)\not\in U^{\bot}\}$, $\chi(g[B])=\{i\in\Omega\mid \eta(i)\not\in U^{\bot}\}$, which, together with (2.6), implies that
\begin{equation}\Wt(f[B])=\Wt(f[X])-\left(\sum_{(i\in\chi(f[X]),\tau(i)\in U^{\bot})}\omega(i)\right),\end{equation}
\begin{equation}\Wt(g[B])=\Wt(g[X])-\left(\sum_{(i\in\chi(g[X]),\eta(i)\in U^{\bot})}\omega(i)\right).\end{equation}
We note that if $f$ and $g$ are locally $\omega$-equivalent, then $\Wt(f[X])=\Wt(g[X])$; moreover, summing both sides of (2.9) over all $I\leqslant_{\mathbb{F}}\mathbb{F}^{[k]}$ with $\dim_{\mathbb{F}}(I)=1$ also implies that $\Wt(f[X])=\Wt(g[X])$. Hence without loss of generality, we assume that $\Wt(f[X])=\Wt(g[X])$ and $k\geqslant2$. Then, by (1), (3.1) and (3.2), $f$ and $g$ are locally $\omega$-equivalent if and only if $\Wt(f[B])=\Wt(g[B])$ for all $B\leqslant_{\mathbb{F}}X$ with $\dim_{\mathbb{F}}(B)=k-1$, if and only if for any $U\leqslant_{\mathbb{F}}\mathbb{F}^{k}$ with $\dim_{\mathbb{F}}(U)=k-1$, (2.9) holds true for $I=U^{\bot}$, which further establishes the desired result.
\end{proof}

\setlength{\parindent}{2em}
For further discussion, we recall the characterization of $\omega$-weight isometries. The following lemma is a special case of either [6, Theorem 5] or [12, Theorem III.1].

\setlength{\parindent}{0em}
\begin{lemma} Let $\varphi\in\End_{\mathbb{F}}(\mathbf{H})$. Then, $\varphi$ is an $\omega$-weight isometry of $\mathbf{H}$ if and only if there exists a bijection $\lambda$ of $\Omega$ such that $\omega(i)=\omega(\lambda(i))$ for all $i\in\Omega$ and $\supp(\varphi(\alpha))=\lambda[\supp(\alpha)]$ for all $\alpha\in\mathbf{H}$.
\end{lemma}

\setlength{\parindent}{2em}

Now we prove Theorem 2.2 with the help of Theorem 2.1 and Lemma 3.1.

\setlength{\parindent}{0em}
\begin{proof}[Proof of Theorem 2.2]
Following the notations in Theorem 2.1, for any $I\leqslant_{\mathbb{F}}\mathbb{F}^{[k]}$ with $\dim_{\mathbb{F}}(I)=1$, (2.9) holds true by Theorem 2.1, which, together with the UDP assumption, further implies that
$$\forall~b\in\mathbb{R}:|\{i\in\chi(f[X])\mid\tau(i)\in I,\omega(i)=b\}|=|\{i\in\chi(g[X])\mid\eta(i)\in I,\omega(i)=b\}|.$$
Hence there exists a bijection $\lambda$ of $\Omega$ and a tuple of non-zero elements $(c_i\mid i\in\Omega)$ in $\mathbb{F}$ such that $\omega(i)=\omega(\lambda(i))$, $\eta(\lambda(i))=\tau(i)\cdot c_i$ for all $i\in\Omega$. Define $Q\in\mathbb{F}^{\Omega\times\Omega}$ as $Q_{i,\lambda(i)}=c_i$ for all $i\in\Omega$, and $Q_{i,j}=0$ for all $j\neq\lambda(i)$; moreover, define $\varphi\in\End_{\mathbb{F}}(\mathbf{H})$ as $\varphi(\alpha)=\alpha Q$. Then, we have $M=LQ$, which implies that $g=\varphi\circ f$; moreover, by Lemma 3.1, $\varphi$ is an $\omega$-weight isometry, as desired.
\end{proof}

\setlength{\parindent}{2em}
Now we are ready to prove Theorem 2.3.

\setlength{\parindent}{2em}
\begin{proof}[Proof of Theorem 2.3]
First, if $\mathbf{H}$ satisfies MEP, then for $\alpha,\beta\in\mathbf{H}$ with $\wt(\alpha)=\wt(\beta)$, the $\omega$-weight preserving map $f\in\Hom_{\mathbb{F}}(\mathbb{F}\alpha,\mathbf{H})$ defined as $f(\alpha)=\beta$ extends to an $\omega$-weight isometry $\varphi$, which further implies that $\beta=\varphi(\alpha)$, as desired.

Second, suppose that $\mathbf{H}$ is transitive. Let $I,J\subseteq\Omega$ with $\sum_{i\in I}\omega(i)=\sum_{j\in J}\omega(j)$. Then, we can choose $\alpha,\beta\in\mathbf{H}$ such that $\supp(\alpha)=I$, $\supp(\beta)=J$. Noticing that $\wt(\alpha)=\wt(\beta)$, we can choose an $\omega$-weight isometry $\varphi\in\End_{\mathbb{F}}(\mathbf{H})$ with $\beta=\varphi(\alpha)$. By Lemma 3.1, there exists a bijection $\lambda:\Omega\longrightarrow\Omega$ such that $\omega(i)=\omega(\lambda(i))$ for all $i\in\Omega$, and $\supp(\varphi(\theta))=\lambda[\supp(\theta)]$ for all $\theta\in\mathbf{H}$. Therefore we have $J=\lambda[I]$, which implies that $|\{i\in I\mid\omega(i)=b\}|=|\{j\in J\mid\omega(j)=b\}|$ for all $b\in\mathbb{R}$, as desired.

Finally, if $(\Omega,\omega)$ satisfies UDP, then for any $C\leqslant_{\mathbb{F}}\mathbf{H}$ and $\omega$-weight preserving map $g\in\Hom_{\mathbb{F}}(C,\mathbf{H})$, by Theorem 2.2, there exists an $\omega$-weight isometry $\varphi\in\End_{\mathbb{F}}(\mathbf{H})$ with $g=\varphi\circ \id_C=\varphi\mid_C$, as desired.
\end{proof}

\setlength{\parindent}{2em}
From now on, we turn to constant weight code. We first use Theorem 2.1 to establish Proposition 2.2. We note that Proposition 2.2 can be proved in a similar way as Theorem 2.1. However, an application of Theorem 2.1 yields the following short proof.

\begin{proof}[Proof of Proposition 2.2]
For the first assertion, let $D,Q\leqslant_{\mathbb{F}}C$ with $\dim_{\mathbb{F}}(Q)=\dim_{\mathbb{F}}(D)$. Then, we can choose $f\in\Aut_{\mathbb{F}}(C)$ such that $f[D]=Q$. Noticing that $\wt(\alpha)=\wt(f(\alpha))$ for all $\alpha\in C$, by Theorem 2.1, we have $\Wt(D)=\Wt(f[D])=\Wt(Q)$, as desired. For the second assertion, let $\alpha,\beta\in C-\{0\}$. Then, we can choose $f\in\Aut_{\mathbb{F}}(C)$ such that $f(\alpha)=\beta$. Noticing that $\Wt(D)=\Wt(f[D])$ for all $D\leqslant_{\mathbb{F}}C$ with $\dim_{\mathbb{F}}(D)=m$, by Theorem 2.1, we have $\wt(\alpha)=\wt(f(\alpha))=\wt(\beta)$, as desired.
\end{proof}

Finally, we prove Theorem 2.4 with the help of Proposition 2.2.

\setlength{\parindent}{0em}
\begin{proof}[Proof of Theorem 2.4]
{\bf{(1)}}\,\,We begin by noting that $\chi(C)=\{i\in\Omega\mid\tau(i)\neq0\}$; moreover, for $B\leqslant_{\mathbb{F}}\mathbb{F}^{k}$ and $D\triangleq\{\gamma G\mid\gamma\in B\}$, we have $\chi(D)=\{i\in\Omega\mid \tau(i)\not\in B^{\bot}\}$ and
\begin{equation}\Wt(D)=\left(\sum_{i\in\chi(C)}\omega(i)\right)-\left(\sum_{(i\in\chi(C),\tau(i)\in B^{\bot})}\omega(i)\right).\end{equation}
Next, if $C$ is a constant weight code, then by Proposition 2.2, all the $(k-1)$-dimensional $\mathbb{F}$-subspaces of $C$ have the same $\omega$-weight, which, together with $\{B^{\bot}\mid B\leqslant_{\mathbb{F}}\mathbb{F}^{k},\dim_{\mathbb{F}}(B)=k-1\}=\{I\leqslant_{\mathbb{F}}\mathbb{F}^{[k]}\mid\dim_{\mathbb{F}}(I)=1\}$ and (3.3), immediately implies (2.10), as desired. Now suppose that (2.10) holds. Then, for $U\leqslant_{\mathbb{F}}\mathbb{F}^{[k]}$, it holds that
\begin{equation}\sum_{(i\in\chi(C),\tau(i)\in U)}\omega(i)=\sum_{(I\leqslant_{\mathbb{F}}U,\dim_{\mathbb{F}}(I)=1)}\sum_{(i\in\chi(C),\tau(i)\in I)}\omega(i)=\frac{(|U|-1)\sigma}{q-1}.\end{equation}
Hence for $D\leqslant_{\mathbb{F}}C$ with $\dim_{\mathbb{F}}(D)=s$ and $B\leqslant_{\mathbb{F}}\mathbb{F}^{k}$ such that $D=\{\gamma G\mid\gamma\in B\}$, from $\dim_{\mathbb{F}}(B^{\bot})=k-s$, (3.3) and (3.4), we deduce that
$$\Wt(D)=\frac{(q^{k}-1)\sigma}{q-1}-\frac{(q^{k-s}-1)\sigma}{q-1}=\frac{(q^{k}-q^{k-s})\sigma}{q-1},$$
as desired.

{\bf{(2)}}\,\,With (1), the desired result follows from the fact that (2.12) always implies (2.10), and conversely, with the UDP assumption, (2.10) implies (2.12).
\end{proof}

\end{document}